\documentclass[12pt]{iopart}

\usepackage{url} 
\usepackage{soul}

\usepackage{upgreek}
\usepackage{graphicx}

\begin{document}

\title[HSC supplies DO to BoB OMZ]{Arabian Sea high salinity core supplies oxygen to Bay of Bengal oxygen minimum zone}

\author{Anoop A. Nayak\textsuperscript{\dag}, P. N. Vinayachandran\textsuperscript{\dag}, and Jenson V. George\textsuperscript{\dag, \ddag(current affiliation)}}
\address{\textsuperscript{\dag}Centre for Atmospheric and Oceanic Sciences, IISc, Bengaluru-560012}
\address{\textsuperscript{\ddag}National Centre for Polar and Ocean Research, Goa, India-403804}

\ead{vinay@iisc.ac.in}

\begin{abstract}
The oxygen minimum zone (OMZ) in the Bay of Bengal (BoB) is unique owing to its curious capability to maintain steady dissolved oxygen (DO) levels. In this study, we identify a process by which the oxygen levels in BoB are sustained above the tipping point, using DO and microstructure profiles in the southern BoB and Argo profiles over the entire basin. High salinity core (HSC) rich in DO is advected by the Summer Monsoon Current (SMC) into BoB. Vertical mixing driven by turbulent and salt-fingering processes recharge DO concentration in thermocline above OMZ. HSC identified in the Argo data, also rich in oxygen, can be traced up to 19$^\circ$ N, confirming that HSC is a source of DO and potentially prevents OMZ from moving to the denitrification regime. In changing climate conditions, this might be the only significant oxygen source for the BoB OMZ in the future.
\end{abstract}

%
\vspace{0.5pc}
%
%
%
%

\section{Introduction}
Oxygen minimum zone (OMZ) is region in the ocean interior where consumption of DO is greater than supply \cite{paulmier2009oxygen}. Mixing oxygen from the atmosphere and photosynthesis by phytoplankton modulate near-surface concentrations of DO. Below the surface, large-scale ocean circulation and vertical mixing are dominant physical processes that control oxygen levels \cite{wyrtki1962oxygen,sarma2002evaluation,jain2017evidence}. The structure of OMZ in the world oceans has two primary controlling factors. Circulation in basins determines the extent of OMZ, whereas biological processes inside OMZ determine oxygen levels \cite{sverdrup1938explanation, wyrtki1962oxygen, vinayachandran2021reviews}. The northern basins of the Indian ocean - the Arabian Sea and BoB host some of the biggest OMZs amongst world oceans \cite{laffoley2019ocean, rixen2020reviews}. BoB OMZ is the fourth largest in world ocean basins \cite{johnson2019oxygen}, with lowest DO levels at the northern edge \cite{d2020structure}.

Oxygen level within OMZs in world oceans is decreasing (deoxygenation) due to global warming \cite{keeling2002change, oschlies2018drivers, breitburg2018declining}. Deoxygenation intensifies denitrification and results in an increase in nitrogen release \cite{stramma2008expanding}. Deoxygenation trends in BoB are weak \cite{brandt2015role, laffoley2019ocean, lachkar2020climatic}, and unlike other OMZs, BoB shows weak signatures of active denitrification \cite{howell1997geochemical, bristow2017n2}. DO levels within BoB OMZ are above threshold, below which prolonged DO values will trigger large-scale denitrification \cite{bristow2017n2, johnson2019oxygen}. DO is suboxic in BoB OMZ but close enough to the threshold that its further decrease can initiate large-scale denitrification \cite{bristow2017n2, d2020structure}. The sources of oxygen or the processes by which oxygen levels are maintained above the tipping point are still to be identified. 


Supply of oxygenated water by eddies has been proposed as a plausible mechanism that prevents denitrification in the BoB \cite{sarma2018ventilation, johnson2019oxygen}. However, several studies suggest the flow of oxygenated water masses as an efficient mechanism. Using model experiments, it was suggested that the oxygen supply to layers above OMZ is essential for maintaining the DO concentrations within OMZ \cite{mccreary2013dynamics}. The Summer monsoon current (SMC) has been well identified as the main pathway for transporting Arabian Sea (AS) high salinity water to the BoB during boreal summer \cite{vinayachandran1999intrusion, jensen2001arabian, webber2018dynamics}. AS water forms the near surface water over AS basin and subducts below the warm, fresh and hence near surface BoB water as it enters BoB. AS water, advected by SMC to the southern BoB mixes with the ambient BoB waters as it moves northward; there is evidence of AS water as north as 12$^\circ$ N \cite{gordon2017intrathermocline, jain2017evidence}. Evidence for oxygenation by Persian Gulf water, advected by SMC, has been presented by \cite{sheehan2020injection}. Although the flow of oxygen-rich water along with SMC in southern BoB has been identified as the essential conduit for the supply chain, the process of oxygenation of BoB water has not been identified.

In this study, we used {\em in situ} oxygen and microstructure profiles to show fluxes of DO from the high salinity core (HSC) into ambient water in the southern BoB. Argo float profiles across the BoB basin are used to understand the spread of high salinity water and its implication on the distribution of DO. Section 2 describes the data used to estimate turbulent fluxes for DO (Detailed descriptions of methods are given in the supporting information). Section 3 presents the DO features and turbulent fluxes observed across the section, followed by results from the set of selected Argo profiles in BoB. Features and statistics estimated from the observations are used to obtain a rough estimate of DO transported in BoB during the summer monsoon. Section 4 discusses summary and conclusions. 

\section{Data and Methods}
{\em In situ} profiles were collected during the Bay of Bengal Boundary Layer Experiment (BoBBLE) from 25 June 2016 to 24 July 2016 \cite{vinayachandran2018bobble}. The measurements included a time series station (TSE) from 4 July 2016 to 14 July 2016 at 89$^\circ$ E, 8$^\circ$ N. A section was occupied twice, along 8$^\circ$ N from 85.3$^\circ$ E to 89$^\circ$ E. Continuous current measurements were carried out using a hull-mounted 150 kHz Teledyne RDI Ocean Surveyor Acoustic Doppler Current Profiler or ADCP (Text S1 of Supporting information). A factory-calibrated SeaBird Electronics (SBE) 9/11 $+$ Conductivity-Temperature-Depth profiler (CTD) was used for measuring vertical profiles of temperature, salinity, and DO (Text S2). Isothermal layer depth (ILD) is defined as the depth where there is a 0.8 $^\circ$C decrease in temperature compared to sea surface temperature \cite{george2019mechanisms}. The depth below which density exceeds near-surface value by an increment due to 0.8 $^\circ$C reduction in temperature is considered the mixed layer depth or MLD \cite{kara2000mixed}. 

Microstructure profiles were collected at CTD stations using a vertical microstructure profiler (VMP - make Rockland Scientific). 2--3 microstructure profiles were collected within an hour during a single VMP operation. VMP operations were roughly at 5.5 hours intervals during the time series. Coefficients of turbulent diffusivity of density ($K_\rho$) and heat ($K_T$) were obtained from the estimates of dissipation rates ($\varepsilon$, $\chi$, Text S3), background stratification and background temperature gradient \cite{osborn1972oceanic, gregg1973microstructure, osborn1980estimates} using the expressions: 

\begin{equation}
	K_\rho = 0.2\frac{\varepsilon}{N^2}; \hspace{1cm}
	K_T = \frac{1}{2}\frac{\chi}{(\nabla\bar{T})^2}
	\label{Equ:KrhoKT}
\end{equation}

Where, $N^2$ is Brunt-V{\"a}is{\"a}l{\"a} frequency and $\nabla\bar{T}$ is background gradient in temperature. If simultaneous profiles of oxygen and diapycnal diffusivity values are available, diapycnal oxygen flux ($F_{DO}$) can be estimated using Fick's law of diffusion \cite{sharples2003quantifying, fischer2013diapycnal, brandt2015role} as follows: 

\begin{equation}
	F_{DO} = -\rho K_{oxy} \frac{\partial DO}{\partial z}
	\label{Equ:FluxDO}
\end{equation}

In equation \ref{Equ:FluxDO}, $\rho$ is density of water, $K_{oxy}$ is diapycnal diffusivity of DO. Estimates of $K_{oxy}$ were obtained from VMP. If salt fingering is indicated at any depth, then $K_{oxy} = R_\rho K_T/\gamma $ \cite{st1999contribution, george2021enhanced}. $R_\rho$ and $\gamma$ are density ratio and flux ratio respectively. If salt fingering is not indicated, then $K_{oxy} = K_\rho$ \cite{fischer2013diapycnal}. $K_\rho$ and $K_{T}$ are estimated following \cite{nayak2022turbulence} and details of methods and constants are given in Text S3. The last term in equation \ref{Equ:FluxDO} is the DO gradient obtained from DO profiles. The second and third terms of Equation \ref{Equ:FluxDO} require that they are estimated from simultaneous measurements \cite{fischer2013diapycnal}. In contrast, VMP and CTD operations were not concurrent, and there were differences in times of VMP and closest CTD operations at a station. Profiles of diffusivity and DO were corrected for alignment before using them in Equation \ref{Equ:FluxDO} (Details of calculation of DO flux are given in Text S4). 

To investigate the spread of oxygen-rich high salinity water into the BoB, we have used profiles from 249 Argo floats deployed between 2002-2021 spread across the BoB (\cite{argo2000argo}; \url{https://argo.ucsd.edu, ://www.ocean-ops.org, ://dataselection.euro-argo.eu/}). The Argo floats provide temperature and salinity profiles at 2 m resolution. Since the depth range of HSC is limited to the top 200 m, we have restricted the use of the Argo data in the top 200 m. All profiles from this collection with 20 $^\circ$C $\leq \theta$ (potential temperature) $\leq$ 27 $^\circ$C, and S $\geq35$ psu within the top 200 m were considered for the analysis. These criteria helped identify Argo profiles with temperature-salinity characteristics of HSC captured during BoBBLE (Analysis of Argo profiles in Section 3). Application of this criteria resulted in 2042 profiles to the north of 8$^\circ$ N in the BoB. 

\section{Results}

\begin{figure}[ht!]
	\includegraphics[width=0.9\textwidth]{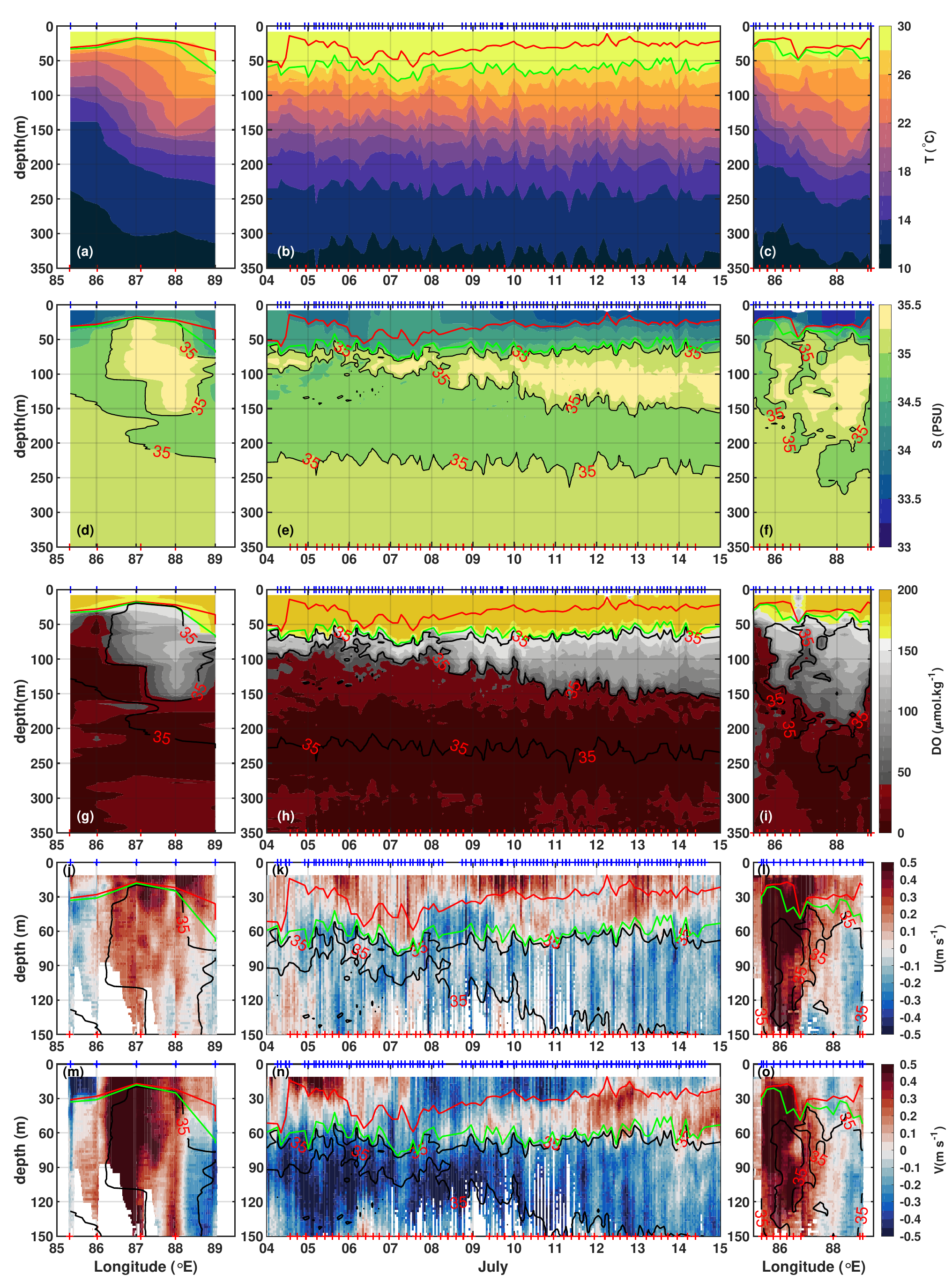}
	\centering
	\caption{BoBBLE observations of temperature (a)--(c), salinity (d)--(f), DO (g)--(i), currents, U (j)--(l) and V (m)--(o). Respective colorbars are to the right of each row. The left and right columns show sections along 8$^\circ$ N executed before and after the time series, respectively. The wide middle column displays the time series at 89$^\circ$ E, 8$^\circ$ N. Blue plus symbol on top of each panel shows the time of CTD operation in the middle column and the location of CTD operation in the left and right columns. Similarly, the red plus symbol at bottom of each panel is for VMP operation. Thick black lines in (d)--(o) are 35 psu contours. Red line is MLD, and green line is ILD in all panels.}
	\label{DO0}
\end{figure}

Stations along the western edge of the 8$^\circ$ N section occupied during BoBBLE were within the seasonal cyclonic circulation called Sri Lanka dome or SLD \cite{vinayachandran1998monsoon}. Temperature contours shoaled to the westernmost stations along the sections, manifesting the impact of SLD (Figure \ref{DO0}a and \ref{DO0}c). TSE was at the eastern edge (Figure S1), and stations between section edges were across the core of SMC \cite{nayak2022turbulence}. SMC, flowing northeastward to the east of SLD, is a source of high salinity Arabian Sea water. This is evident from the stations showing a high salinity core along with strong currents (Figure \ref{DO0}j, \ref{DO0}l, \ref{DO0}m, and \ref{DO0}o). A subsurface high salinity core (HSC, S $>35$ psu, \cite{vinayachandran2013summer}) was present at TSE and at the stations crossing SMC, but not at the stations within SLD. HSC was thickest ($\sim$ 110 m) across the first 8$^\circ$ N section (Figure \ref{DO0}d). At TSE (Figure \ref{DO0}e), HSC thickness increased from ($\sim$ 20 m) at the start of the time series (4--7 July 2016) to $\sim$ 90 m towards the end (14 July 2016).

\subsection{Dissolved oxygen}
Highest DO values ($\sim$ 185 $\upmu$mol kg$^{-1}$) were observed near the surface within ILD (Figure \ref{DO0}g-i). Below ILD, DO contours followed salinity contours in top 200 m. Depths of contours with lower DO ($\sim$ 50 $\upmu$mol kg$^{-1}$) increased with an increase in thickness of HSC in the subsurface \cite{roy2021southern}. DO decreased from 150 $\upmu$mol kg$^{-1}$ at the top of HSC to 50 $\upmu$mol kg$^{-1}$ at the bottom (Figure \ref{DO0}). The thickness of the oxycline ($\sim$ 50--150 $\upmu$mol kg$^{-1}$ layer) was lowest at stations within SLD where HSC was absent. Contours of lower DO ($<$ 50 $\upmu$mol kg$^{-1}$) were observed at depths shallower than 50 m \cite{roy2021southern} at SLD stations while they were deeper than the depth of bottom of HSC at other stations. These figures suggest that HSC is a source of DO in the BoB. 

\subsection{DO fluxes}
\begin{figure}[ht!]
	\includegraphics[width=1.1\textwidth]{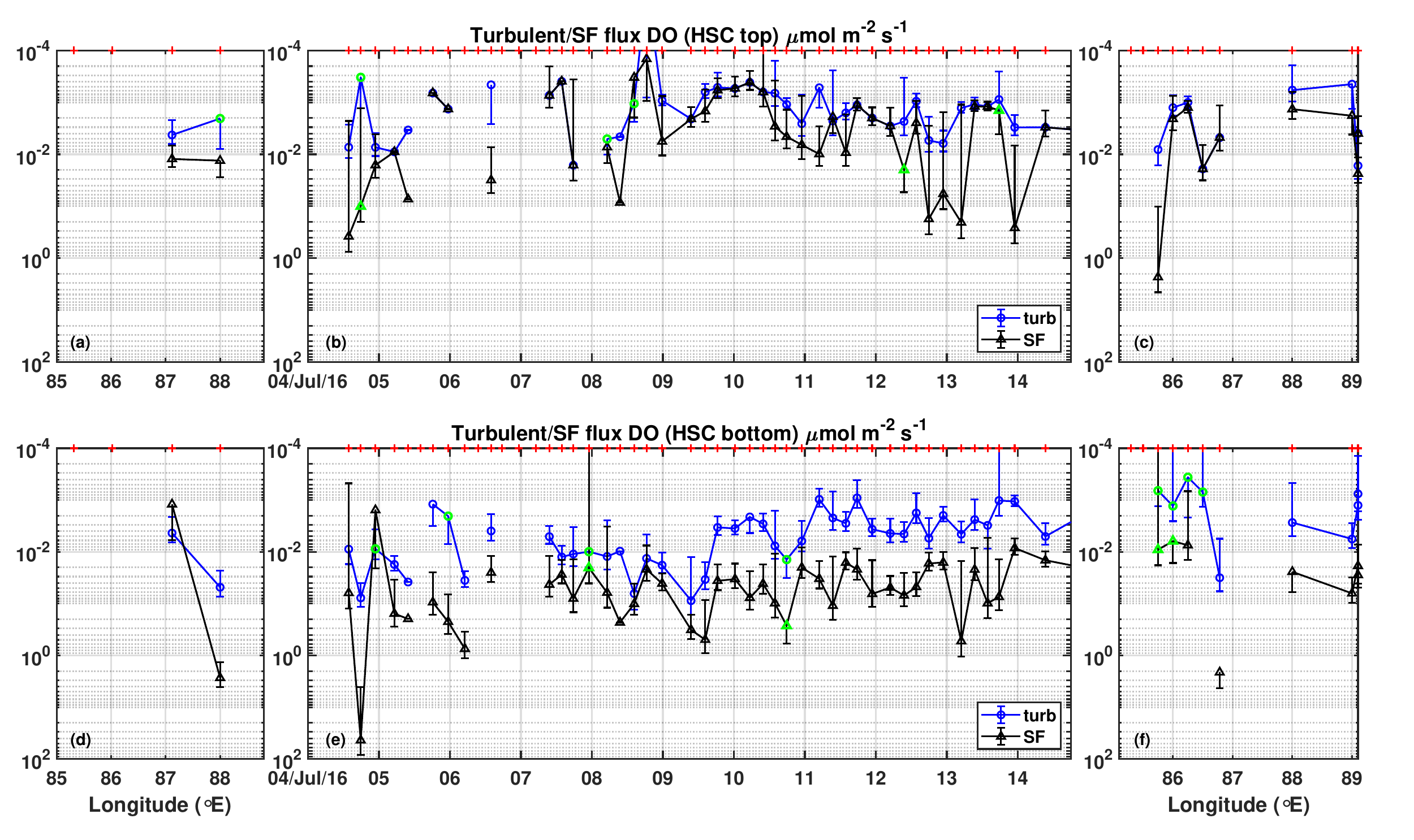}
	\centering
	\caption{Layer-average DO fluxes in the downward direction. The columns are arranged as in Figure 1. $turb$ and $SF$ are layer average turbulent (blue) and salt fingering (black) fluxes, respectively. (a--c) Fluxes for top layer of HSC. (d--f) Fluxes for bottom layer of HSC. The bottom layer shows higher fluxes than the top layer. Green markers are for layer-average fluxes directed upward. Fluxes are shown for profiles with HSC. Error bars are standard errors.}
	\label{DO5_2}
\end{figure}

DO fluxes were generally directed downward since the DO values decreased with depth. DO fluxes from figure S4 were averaged along the vertical within HSC in two layers to distinguish between processes at the top and bottom of HSC (Text S5). Figure \ref{DO5_2} shows that the bottom layer average turbulent fluxes were of O(10$^{-3}$--10$^{-2}$ $\upmu$mol m$^{-2}$ s$^{-1}$) in time series and in sections. The top layer averages show turbulent fluxes of O(10$^{-4}$--10$^{-3}$ $\upmu$mol m$^{-2}$ s$^{-1}$) from 9 July to the end of the time series, lower than fluxes in bottom layer. Stations within SLD show weak downward DO fluxes below the mixed layer and upward turbulent fluxes above the 35 psu contour (first profile in the first column of figure S4). A profile within SLD had a lower value of oxygen minima than other stations, which suggests an effect of the SLD cyclonic circulation on DO values in the subsurface (Figure \ref{DO0}g). 

Salt fingering processes enhance the fluxes below the HSC \cite{george2021enhanced}. The salt-fingering DO fluxes for the bottom layer were 1-1.5 orders higher than the layer average turbulent fluxes (Figure \ref{DO5_2}). Average DO fluxes in the bottom layer were higher than in the top layer, suggesting that HSC loses oxygen to the subsurface more than it receives at its top from the mixed layer. The average fluxes are high in the bottom layer because of more events of higher-order diffusivity values than in the upper layer. Two orders difference between the mean and median of $K_{SF}$ values within the bottom layer of HSC (Figure S2) suggest that high flux processes are intermittent. To summarize, at SLD stations there is a loss of DO in deeper layers, and at the stations with HSC there is a loss of DO to deeper layers. 

\subsection{Spread of HSC using Argo Data}
\begin{figure}[ht!]
	\includegraphics[width=\textwidth]{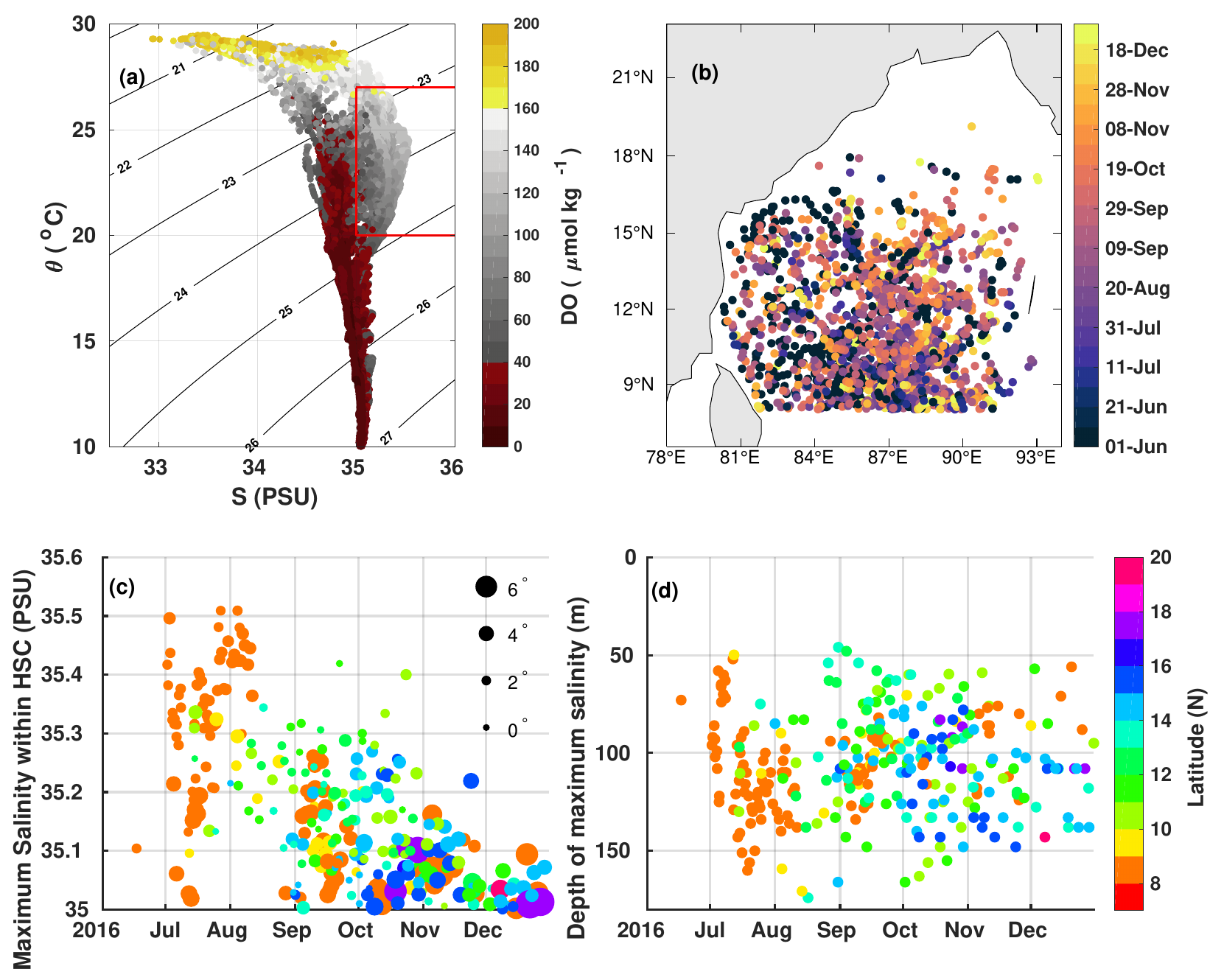}
	\centering
	\caption{(a) $\theta$-S diagram from CTD observations in BoBBLE. The colorbar shows DO. Box wraps points having S $>$ 35 psu and (20 $^\circ$C $<$ $\theta$ $<$ 27 $^\circ$C). Points to the higher side of 35 psu have higher DO than those on the lower side. This shows that high salinity water carries higher DO. (b) Locations of Argo profiles with HSC signature. Near the 8$^\circ$ N latitude, profiles with HSC signatures were observed around the end of the year. Few profiles in the northern BoB (as far as 19$^\circ$ N) were obtained at the end of the corresponding year. (c) Y-axis scales the maximum salinity value within the HSC found in the selected Argo profiles from June-December 2016. The sizes of bullets are proportional to the distance (in degrees) of profile location from 85$^\circ$ E longitude (Black bullets provide sizes for distances in degrees). Maximum salinity decreased with an increase in latitude and time. (d) Y-axis shows the depth of maximum salinity varying between 80--100 m. The color in (c and d) shows latitude.}
	\label{DO6}
\end{figure}

Argo profiles having HSC signatures are densely concentrated in the southern BoB (mostly between 86--88$^\circ$ E, Figure \ref{DO6}). Figure \ref{DO6}c, \ref{DO6}d show locations and times of Argo profiles selected for year 2016. The size of bullets increases with time in figure \ref{DO6}c, suggesting that HSC signatures spread through the BoB after entering the basin. The colors indicate a northward movement of HSC. Depths of maximum salinity of HSC identified at respective Argo locations (Figure \ref{DO6}d) were within the range 50--150 m, also shown in \cite{gordon2016bay}. These depths are close to values shown in model simulations by \cite{singh2022front}. Argo profiles suggest that high salinity Arabian Sea water enters the southern BoB and spreads in BoB along these depths. The decrease of salinity values in figure \ref{DO6}c suggests HSC mixing with ambient water as it moves northward. 

\begin{figure}[ht!]
	\includegraphics[width=\textwidth]{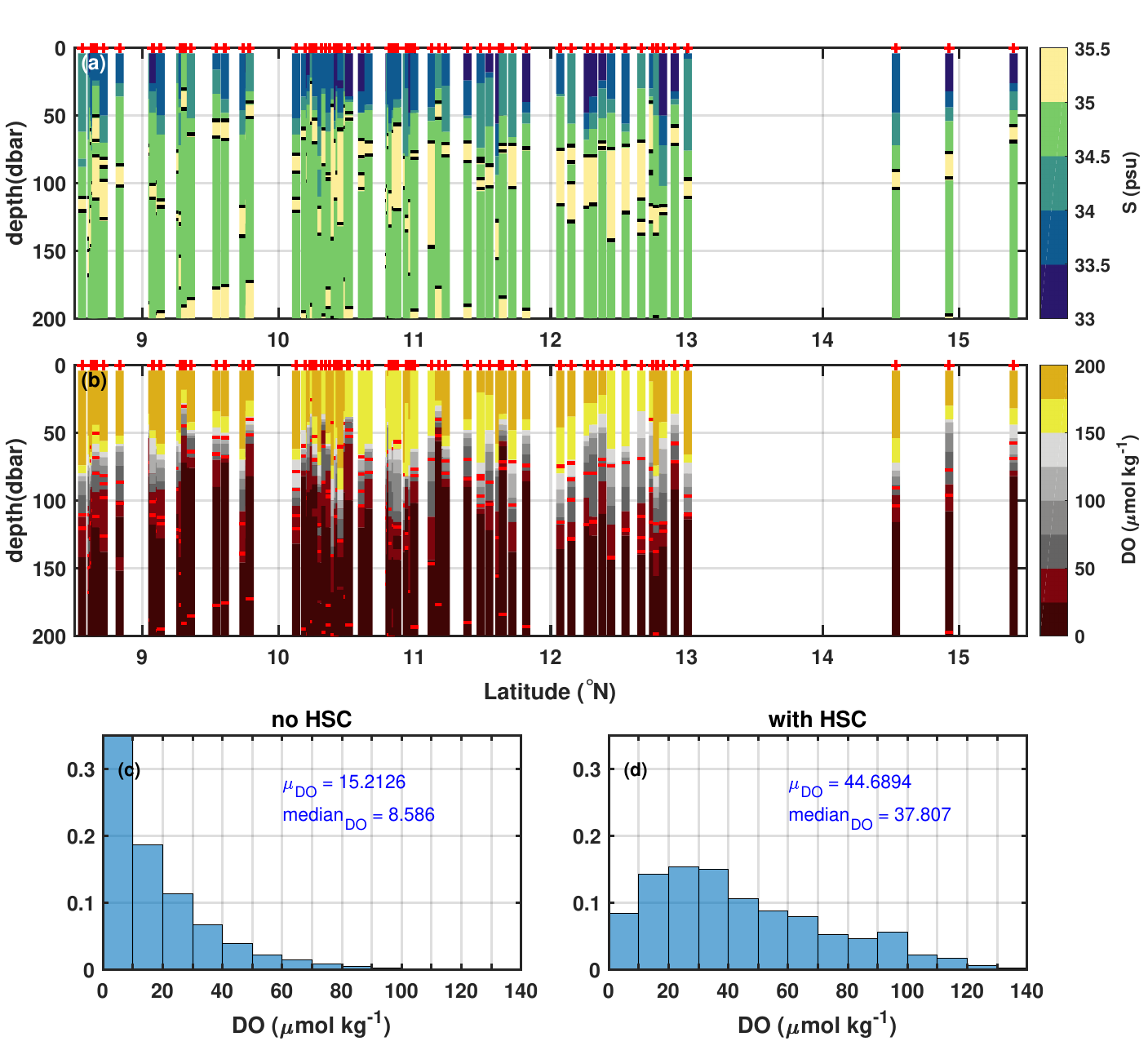}
	\centering
	\caption{South-north section of Argo profiles which had both (a) salinity and (b) DO using 79 profiles discussed in section 3.3. Red plus signs on top of (a) and (b) mark latitudes where the profiles were located. 35 psu salinity contours are marked black in (a) and red in (b). Comparison of normalized histogram of DO from Argo profiles identified (d) \textbf{with HSC} to DO from Argo profiles identified with (c) \textbf{no HSC}. The peak of the histogram \textbf{with HSC} has a higher DO than the peak of the histogram with \textbf{no HSC}. Mean and median DO are mentioned in blue.} 
	\label{D10}
\end{figure}

A total of 79 Argo profiles with high salinity signatures north of 8$^\circ$ N between June-December of 2002-2021 had both salinity and DO profiles. Figure \ref{D10} shows a south-north section of salinity and DO from the 79 profiles. Figure \ref{D10} shows higher concentration of profiles to the south of 13$^\circ$ N than to the north. Sudden jumps in salinity and DO contours are because the profiles are identified at different parts of the basin at different times. Since the Argo profiles are limited in number and non-uniform in basin coverage, tracking high salinity signatures is unreliable. However, figure \ref{D10}a reveals a general picture that HSC is present at depths shallower than 150 m up to 15$^\circ$ N. Most of these depths with high salinity signatures showed deeper oxycline, similar to results discussed in the BoBBLE observations (Figure \ref{DO0}g-i). 

Figure \ref{DO6}a shows that the high salinity water found in BoBBLE observations inside the box (based on $\theta$ and salinity criteria in section 2) is within isopycnals 22.7--24.9 kg m$^{-3}$ (Box shown in Figure \ref{DO6}a). DO values from the Argo profiles located within these isopycnals are separated in two categories based on the salinity criterion (Figure \ref{D10}c, \ref{D10}d). If the DO values have corresponding salinity $>$ 35 psu then they are considered points \textit{with HSC}. Else DO values are considered as points \textit{without HSC}. 7278 points were selected from 1382 profiles \textit{without HSC} and 595 points were selected from 79 profiles \textit{with HSC}. The mean and median DO from points \textit{with HSC} were higher than the mean and median DO from points \textit{without HSC}. The points \textit{without HSC} are skewed to lower DO values (Figure \ref{D10}c). This difference in DO between \textit{with HSC} and \textit{without HSC} is the excess DO that HSC adds to the BoB. Oxygen rich high salinity water spreads in BoB and mixes with BoB water between these isopycnals \cite{jain2017evidence}. The resultant water mixture between these isopycnals will be rich in DO. DO generally decreases from high ($\sim$ 180 $\upmu$mol kg$^{-1}$) values near the surface to low ($<$ 50 $\upmu$mol kg$^{-1})$ values below the surface. Figure \ref{D10} suggests that this decreasing trend is slower for water columns with high salinity water. Therefore, high salinity water transports relatively higher DO and replenishes the deeper water with DO.  

\subsection{Transport of DO}
The expression for oxygen supplied by the SMC into the BoB is $Q = (W \times H \times U \times \Delta t) \times (\rho_{AS} \times \Delta DO)$ \cite{sheehan2020injection}. The product of quantities in the first bracket is volume transported by SMC. Here, $W$ is the width of HSC and is taken as 200 km (Figure \ref{DO0}). $H$ is the thickness of HSC, which loses salinity and DO to the bottom layers. We have taken $H \approx$ 50 m. Current speed, $U \approx$ 0.5 m s$^{-1}$ after considering currents within the top 150-200 m (Figure \ref{DO0}j-\ref{DO0}o). Assuming the effective transport of HSC at the southern BoB lasts for a month, $\Delta t \approx$ 30 days. The product in the second bracket is the excess DO per unit volume transported by SMC. Density ($\rho_{AS}$) $\approx$ 1023.5 kg m$^{-3}$ since the Argo profiles were considered based on the presence or absence of HSC within a layer between the isopycnals 22.7 -- 24.9 kg m$^{-3}$. $\Delta$DO is difference of DO values between points \textit{with HSC} and \textit{without HSC}. Figure \ref{D10} shows that the difference in median values of DO between points \textit{with HSC} and \textit{without HSC} is more than 25 $\upmu$mol kg$^{-1}$. For the estimate here, we have assumed $\Delta$DO $=$ 20 $\upmu$mol kg$^{-1}$. This is to the lower side of the difference of medians, considering the bias due to the high density of floats identified in the southern BoB. These values give an estimate of $Q = 2.65 \times 10^{17} \upmu$mol.

The estimate $Q$ is the excess DO that SMC transports to the layer between the isopycnals mentioned. Assuming the basin size of 1000 km $\times$ 1000 km and the layer thickness as 120 m, the total increase in the DO is 2.15 $\upmu$mol kg$^{-1}$. The actual basin area covered by the high salinity water is lesser depending on the circulation within the basin \cite{sheehan2020injection}. The estimate $Q$ is an order higher than that estimated by \cite{sheehan2020injection} because the AS water has a larger volume than the Persian Gulf water \cite{jain2017evidence, sheehan2020injection}. Decrease in DO due to oxygen consumption rate in subsurface ($\sim$ 81 Tg yr$^{-1}$) is around 1.72 $\upmu$mol kg$^{-1}$ in 30 day period \cite{sarma2002evaluation}. 

The contribution of AS water to BoB OMZ will depend on the vertical mixing processes encountered by the northward-moving oxygenated HSC. Section 3.2 showed that net DO fluxes out of HSC due to turbulence is of O(10$^{-3}$--10$^{-2}$ $\upmu$mol m$^{-2}$ s$^{-1}$) and that due to salt fingering is of O(10$^{-2}$--10$^{-1}$ $\upmu$mol m$^{-2}$ s$^{-1}$). The contribution of DO fluxes is $\sim$ 0.01--0.1 $\upmu$mol kg$^{-1}$. There is the seasonality in subsurface mixing within BoB \cite{cherian2020seasonal}. The vertical fluxes are weak in July and two orders higher in September-October than in July. This agrees with the increase in contribution of vertical fluxes to OMZ towards the end of year discussed in \cite{sarma2002evaluation}. Moreover, the BoB receives high vertical organic matter fluxes due to river discharge increasing oxygen demand in subsurface layers \cite{ittekkot1991enhanced, prakash2013can}. Thus, the contribution of the HSC in recharging the subsurface OMZ with DO depends on the variation in the distribution of high salinity water, vertical processes, and productivity in the BoB basin. 

\section{Summary and Conclusions}

\noindent Processes that can supply DO to BoB OMZ are of great interest owing to curious capacity of the OMZ to stay above the tipping point. Due to strong stratification in the northern BoB, recharging DO in the subsurface layers via air-sea interaction  is highly improbable or extremely inefficient \cite{d2020structure}. Cyclonic eddies intensify OMZ in BoB through an increase in organic matter production, and anticyclonic eddies weaken the OMZ through ventilation of DO into the OMZ in BoB \cite{sarma2018role, sarma2018ventilation}. Relative number of cyclonic and anticyclonic eddies affect the balance of DO within the OMZ. A more significant factor in controlling DO concentrations in the OMZ is the oxygen supply through the thermocline via sinking processes \cite{sarma2002evaluation}. Replenishing layers above BoB OMZ with DO is important for maintaining DO within OMZ \cite{mccreary2013dynamics}. 

During the summer monsoon, SMC advects high salinity water from the Arabian Sea into the BoB. Observations across the SMC in the southern BoB show that subsurface high salinity core is rich in DO (Figure \ref{DO0}). Figure \ref{DO6}b shows locations of Argo profiles throughout the BoB that carry TS signatures of HSC observed in the southern BoB. Salinity maxima within HSC are highest at the southern latitudes of BoB during July and decrease with an increase in time and latitude (Figure \ref{DO6}c). These results suggest that HSC enters the southern BoB, moves northwards, and diffuses its salinity to the ambient lower salinity BoB water. HSC can be traced to the northern parts of the BoB \cite{singh2022front}, consistent with inferences from observations in this study. Estimates of turbulent and salt fingering fluxes for DO show that the HSC loses DO to layers below due to small-scale processes. The turbulent DO fluxes estimated in this study are comparable to average fluxes presented in \cite{fischer2013diapycnal, brandt2015role}, but these studies did not consider salt fingering processes. Our analyses of {\em in situ} profiles show that the turbulent fluxes are important to recharge the deeper BoB layers with HSC as a source of oxygen-rich water.  

Results presented here are significant in view of the long-term changes likely to manifest in the BoB. Some studies have suggested that streams like the Persian Gulf, which transports DO at depths of BoB OMZ \cite{jain2017evidence, sheehan2020injection}, will weaken due to climate change leading to the intensification of BoB OMZ \cite{laffoley2019ocean}. Climate change will increase the stratification \cite{yamaguchi2019trend} and further restrict vertical exchange of momentum. Lower vertical momentum exchange will reduce oxygen transfer from mixed layer to deeper waters. Increased stratification will strengthen the barrier layer in the BoB; therefore, surface wind forcing  will be ineffective in mixing DO to depths below the mixed layer. In such a scenario, horizontal transport of DO in subsurface layers will play an essential role in replenishing the DO values in subsurface layers, OMZ. Therefore, it is essential to study impact of SMC on the subsurface DO in BoB using comprehensive observations and model simulations.


\ack
The Ministry of Earth Sciences, the Government of India, funded the BoBBLE field program on board the R/V Sindhu Sadhana under its Monsoon Mission program administered by the Indian Institute of Tropical Meteorology, Pune. PNV acknowledges partial financial support from the J C Bose fellowship provided by SERB, DST, Govt. of India. AAN thanks Nihar Paul for his input in the Argo data analysis. Argo data were collected and made freely available by the International Argo Program and the national programs that contribute to it. (\url{https://argo.ucsd.edu}, \url{https://www.ocean-ops.org}). The Argo Program is part of the Global Ocean Observing System.\\


\noindent\textbf{References}\\

\bibliography{references_GRL}

\end{document}


\title[]{Supplementary Information for ``Arabian Sea high salinity core supplies oxygen to Bay of Bengal oxygen minimum zone"}

\author{Anoop A. Nayak\textsuperscript{\dag}, P. N. Vinayachandran\textsuperscript{\dag}, and Jenson V. George\textsuperscript{\dag, \ddag(current affiliation)}}
\address{\textsuperscript{\dag}Centre for Atmospheric and Oceanic Sciences, IISc, Bengaluru-560012}
\address{\textsuperscript{\ddag}National Centre for Polar and Ocean Research, Goa, India-403804}

\ead{vinay@iisc.ac.in}
\vspace{10pt}



%
%
%
%
%
\vspace{10pt}
\noindent\textbf{Contents of this file}
\begin{enumerate}
	\item Text S1 to S6
	\item Figures S1 to S4
\end{enumerate}

\section*{Introduction}
This supporting information provides details of the data used in the study and methods to calculate the dissolved oxygen fluxes. The fluxes were calculated using two different observation platforms, each contributing to different terms in the flux equation; diffusivity from the Vertical microstructure profiler (VMP) and dissolved oxygen from the CTD platform. Therefore, the method required correction for the alignment of profiles. Procedures to correct the profiles are discussed. A summary of diffusivities and dissolved oxygen fluxes from data are also discussed here. Figures are at the end of this document. 


\section*{Text S1}
\underline{Details of the ADCP measurements:} Continuous current measurements were carried out using a hull-mounted 150 kHz Teledyne RDI Ocean Surveyor Acoustic Doppler Current Profiler (ADCP). The vertical resolution was 2 m, and single ping data were at 8--s intervals. Before calculating gradients of ADCP currents, a 1--hour moving average was applied along time and a 4--m moving average along the vertical direction to reduce the effect of noise. The choice of 4--m was consistent with the 8--s slots used for microstructure estimates.

\section*{Text S2}
\underline{Details of the dissolved oxygen measurements:} A factory-calibrated SeaBird Electronics (SBE) 9/11 $+$ Conductivity-Temperature-Depth profiler (CTD) was used for measuring vertical profiles of temperature, salinity, and dissolved oxygen. 12 Niskin Sampler bottles of 10 L each on a Rosette Frame were used for collecting water samples at discrete depths for measuring various biogeochemical parameters. Analysis of dissolved oxygen (DO) at a precision of $\pm$ 0.03 $\upmu$M was based on the \cite{carpenter1965accuracy} modification of the traditional Winkler titration. These DO data sets were used to calibrate the CTD measurements of DO. There was significant agreement between Winkler DO and CTD DO with coefficient of determination (r = 0.99, n = 306, p $< 0.01$).

\section*{Text S3}
\underline{Details of the VMP measurements:} VMP had two airfoil shear sensors and one FP07 temperature sensor, which measured microstructure shear and temperature gradients with a sampling rate of 512 Hz. CT sensors (JFE Advantech, sampling rate of 64 Hz) measured accurate temperature and salinity. The top 10 m of the microstructure observations were ignored to avoid contamination from ship noise. The shear and temperature gradient microstructure from VMP were used to estimate dissipation rates of turbulent kinetic energy ($\varepsilon$) and temperature variance ($\chi$) \cite{bluteau2011estimating, bluteau2017determining}. The shear and temperature gradient microstructure profiles were segmented (8--s; approximately 4--m of data for average fall speed of VMP $\sim$ 0.5 m s$^{-1}$) in the vertical, and each segment is used to calculate spectra of the temperature gradient. Integral and spectral fitting methods were applied over each segment of temperature gradient microstructure to estimate $\chi$ corresponding to $\varepsilon$ \cite{bluteau2017determining}.

Coefficients of turbulent diffusivity of density ($K_\rho$) and heat ($K_T$) were obtained from the estimates of dissipation rates ($\varepsilon$, $\chi$), background stratification and background temperature gradient \cite{osborn1972oceanic, gregg1973microstructure, osborn1980estimates}. The expressions are as follows:

\begin{equation}
	K_\rho = 0.2\frac{\varepsilon}{N^2}; \hspace{1cm}
	K_T = \frac{1}{2}\frac{\chi}{(\nabla\bar{T})^2}
	\label{Equ:KrhoKT}
\end{equation}

Where, $N^2$ is Brunt-Vaisala frequency and $\nabla\bar{T}$ is background gradient in temperature. Tracer diffusion from HSC in the southern Bay of Bengal (BoB) to ambient water can occur by turbulent and double-diffusive processes (George et al., 2019, 2021). Since the high salinity core observed across the measurements was warmer, some locations were active in salt fingering, and this process enhanced diffusivity \cite{george2021enhanced}. Turner angle ($Tu$; \cite{ruddick1983practical}) is a measure of identifying possible sites for double-diffusive convection and salt fingering processes. It is calculated as $Tu = tan^{-1}\left(\frac{N^2_T-N^2_S}{N^2_T+N^2_S}\right)$ (in degrees). Here $N^2_T$ and $N^2_S$ are contribution of temperature and salinity to $N^2$ respectively \cite{george2021enhanced}. $Tu$ values in between (45,90) degrees suggest the presence of salt fingering processes. The mixing coefficient ($\Gamma = \left[0.5N^2\chi\right]/\left[\epsilon\left( \frac{d\bar{T}}{dz} \right)^2\right]$; \cite{oakey1985statistics, hamilton1989vertical}) is also considered to distinguish sites for salt fingering \cite{george2021enhanced}. If $\Gamma > 0.3$ and $45<Tu<90$ then eddy diffusivity due to salt fingering is modelled as $K_{SF} = R_\rho K_T/\gamma $ \cite{st1999contribution} where $R_\rho$(density ratio) $ = \alpha(\frac{d\bar{T}}{dz})/\beta(\frac{d\bar{S}}{dz})$ and $\gamma$ (flux ratio) is considered equal to 0.7 \cite{fernandez2014microstructure, george2021enhanced}. This method gives diffusivity profiles at all VMP stations, including turbulent and salt-fingering processes.

\section*{Text S4}
\underline{Details of the estimation of DO flux:} Diapycnal oxygen flux can be estimated using Fick's law of diffusion if simultaneous profiles of oxygen and diapycnal diffusivity values are available \cite{sharples2003quantifying, fischer2013diapycnal, brandt2015role}. The expression for diapycnal oxygen flux ($F_{DO}$) is as follows:

\begin{equation}
	F_{DO} = -\rho K_{oxy} \frac{\partial DO}{\partial z}
	\label{Equ:FluxDO}
\end{equation}

In equation \ref{Equ:FluxDO}, $\rho$ is density of water, $K_{oxy}$ is diapycnal diffusivity of dissolved oxygen. Estimates of $K_{oxy}$ were obtained from VMP; $K_{oxy} = K_\rho$ \cite{fischer2013diapycnal}. $K_{oxy} = K_{SF}$ if salt fingering is indicated at the depth. The last term in equation \ref{Equ:FluxDO} is the gradient of dissolved oxygen obtained from dissolved oxygen profiles measured by the CTD platform. VMP and CTD operations were not concurrent, and there were differences in times of VMP and closest CTD operations at a station, whereas the second and third terms of equation \ref{Equ:FluxDO} required that they were estimated from simultaneous measurements \cite{fischer2013diapycnal}. We have considered profiles of DO closest in time of VMP operation for the estimation of $F_{DO}$. To alleviate the error originating from this disparity, we have aligned the profiles by a vertical shift. This study intends to use diffusivity estimates from VMP and the background gradient of DO from CTD. It is necessary to correct profiles for the displacements because the coefficient of diffusivity is a measure of the microstructure activity at a depth. It also depends on the background gradients of temperature and salinity at the particular depth (Equation \ref{Equ:KrhoKT}). 

Features in CTD temperature and salinity profiles did not overlap and instead were vertically displaced with respect to VMP profiles. We have used isothermal layer depth to determine the relative displacement. Figure 1 shows that the top of HSC and 150-160 $\upmu$mol kg$^{-1}$ DO isoline follow the bottom of the near-surface isothermal layer. We have shifted the selected CTD salinity profile vertically by a depth equal to the difference in ILDs observed at CTD and VMP. Hereafter, shifting of profiles will mean the operation discussed before. DO profiles from CTD were shifted before using it for the second term of Equation 2. Such a shift of profiles will ensure the coefficients of diffusivity and background gradients of DO for equation 2 are for the same feature in vertical. We have compared the corresponding gradients of shifted profiles of salinity with gradients of salinity estimated from salinity profiles by VMP slow sensors. A similar comparison is made using the non-shifted profiles. Figure \ref{DO1p1} shows assuring numbers by increased correlation coefficients and reduced errors for shifted profiles. However, the correlation coefficient is unequal to unity since the CTD and VMP operations were not simultaneous. Therefore, there is a difference in features captured by individual observation platforms.

\section*{Text S5}
\underline{Distribution of the diffusivity values:} Diffusivities ($K_\rho$ and $K_{SF}$) estimated using VMP measurements were divided into two layers within HSC to distinguish the processes at the top and bottom of the HSC. The top layer of HSC extends between the top 35 psu contour and the depth of maximum salinity. The bottom layer extends downwards from the depth of maximum salinity to double the distance between the depth of maximum salinity and the bottom 35 psu contour. The bottom layer is thicker to include the depths with high values of dissipation rates below the 35 psu contour as shown in \cite{george2019mechanisms, george2021enhanced}. The distribution of diffusivities is due to the effect of the background state on the generation of turbulent processes (Figure S\ref{DO4_1}). Due to reduced stratification in the bottom layer of HSC, turbulent events are more likely within the bottom layer of HSC than the top of HSC \cite{nayak2022turbulence}. HSC was warmer than ambient water, which gave way to salt fingering processes at its bottom and enhanced diffusivities \cite{george2021enhanced}.

\section*{Text S6}
\underline{DO fluxes in BoBBLE:} Fluxes were generally directed downward (blue shade; Figure S\ref{DO5}) since the dissolved oxygen values decreased with depth. The bottom panels of Figure S\ref{DO5} show higher fluxes than the top panels at the bottom edge of HSC, suggesting an increased diapycnal flux due to salt fingering. Strong downward turbulent fluxes of O(10$^{-3}$-10$^{-2}$ $\upmu$mol m$^{-2}$ s$^{-1}$) were observed along the 35 psu contour at the bottom edge of HSC. Fluxes of O($>$ 10$^{-2}$ $\upmu$mol m$^{-2}$ s$^{-1}$) were enhanced due to salt fingering during the time series (Figure S\ref{DO5}). Below the 35 psu (top isoline of HSC), within the HSC, fluxes were of O(10$^{-4}$-10$^{-3}$ $\upmu$mol m$^{-2}$ s$^{-1}$) with spots of upwards fluxes (Figure S\ref{DO5}). Stations within SLD show weak downward DO fluxes below the mixed layer and upward turbulent fluxes above the deeper 35 psu contour (first profile in the first column of figure S\ref{DO5}). A profile within SLD shows a lower value of oxygen minima which suggests an effect of the SLD cyclonic circulation on DO values in the subsurface (Figure 1g). The minimum DO found at 35 psu contour of the same profile was lower than all other stations. At the beginning of the time series, the HSC behaved like a thin layer and has points missing in the top panel of figure 2.\\

\noindent\textbf{References}\\

\bibliography{references_GRL}

\begin{figure}[ht!]
	\includegraphics[width=0.8\textwidth]{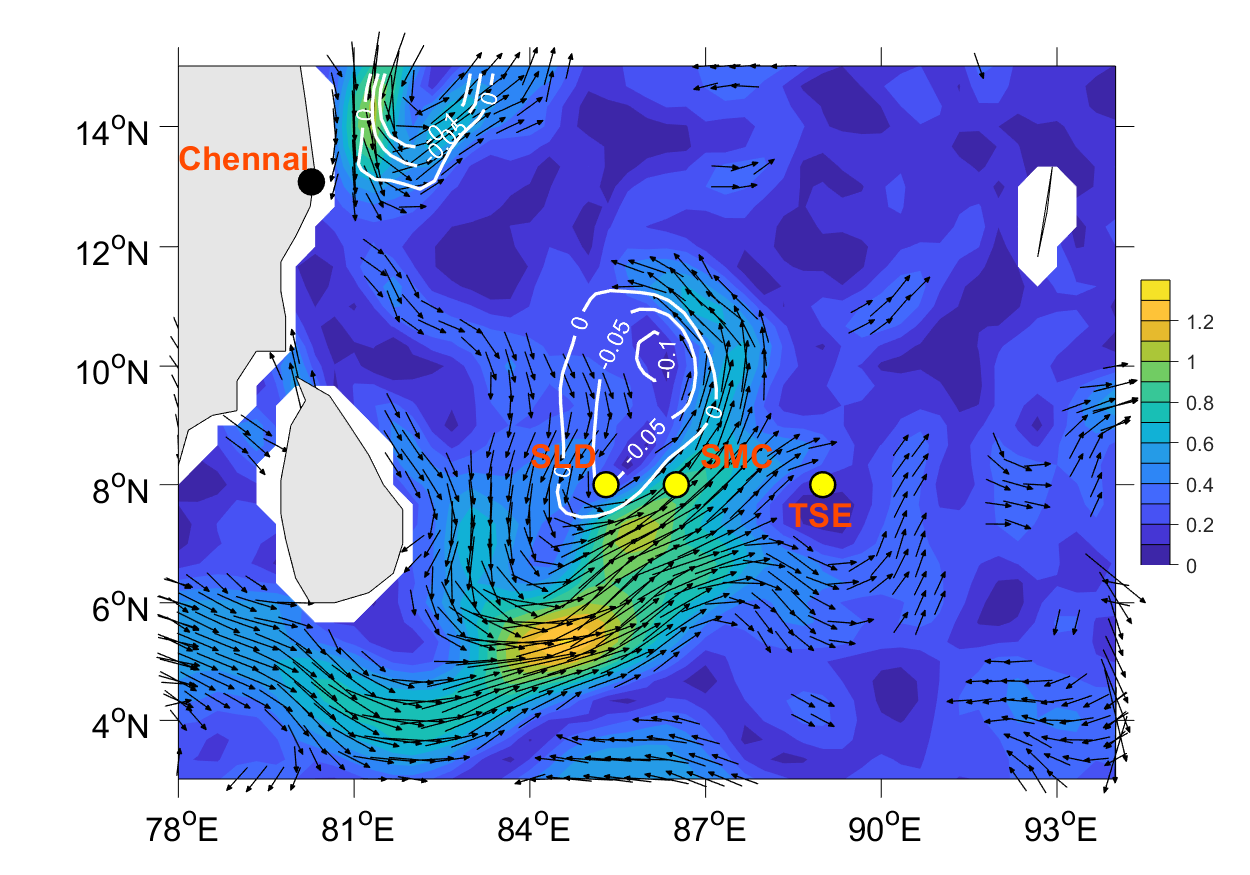}
	\centering
	\caption{Locations of stations within SLD, SMC, and at TSE. SLD stations were the westernmost stations in BoBBLE. A time series station was located at TSE (89$^\circ$ E, 8$^\circ$ N), which was the easternmost of the stations. SMC stations were aligned across the strong northeastward currents. Figure 1 of \cite{nayak2022turbulence} shows the exact locations of several stations counted as SLD and SMC stations during BoBBLE. Chennai was the point of the port. Vectors (for speeds $>$ 0.3 m s$^{-1}$) and color (m s$^{-1}$) are current vectors and magnitude, respectively, at the start of the BoBBLE mission, end of June 2016. White contours show negative Sea Surface Height Anomalies (SSHA, m), manifesting the SLD. Currents from Ocean Surface Currents Analyses Real Time (OSCAR) datasets and SSHA from Copernicus Climate Change Service.}
	\label{DO1_1}
\end{figure}

\begin{figure}[ht!]
	\includegraphics[width=\textwidth]{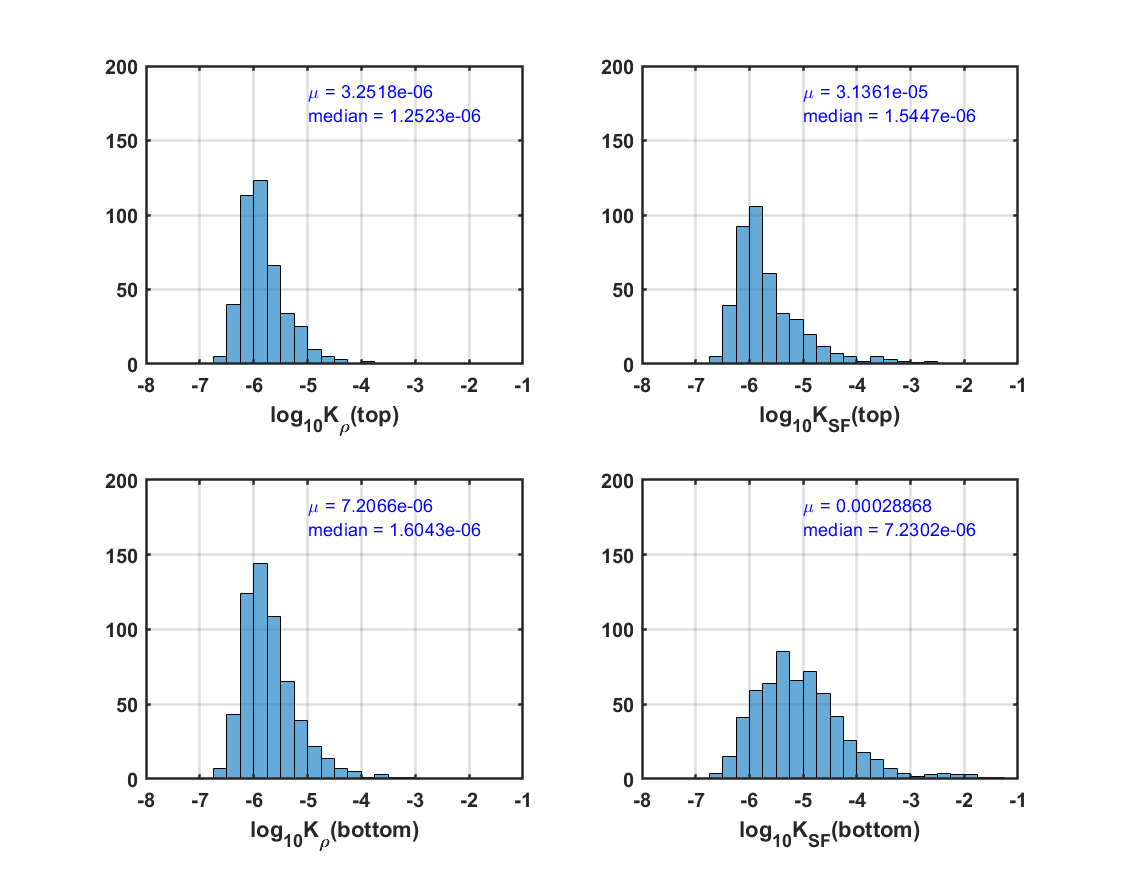}
	\centering
	\caption{The top rows are the distribution of $K_\rho$ (left) and $K_{SF}$ (right) in the top layer of HSC, and the bottom row shows the distribution of $K_\rho$ (left) and $K_{SF}$ (right) in the bottom layer of HSC. The distribution at the bottom right is skewed to higher diffusivity values due to salt fingering. Mean and median values (in m$^2$ s$^{-1}$) are shown at the top right of each window.}
	\label{DO4_1}
\end{figure}

\begin{figure}[ht!]
	\includegraphics[width=0.8\textwidth]{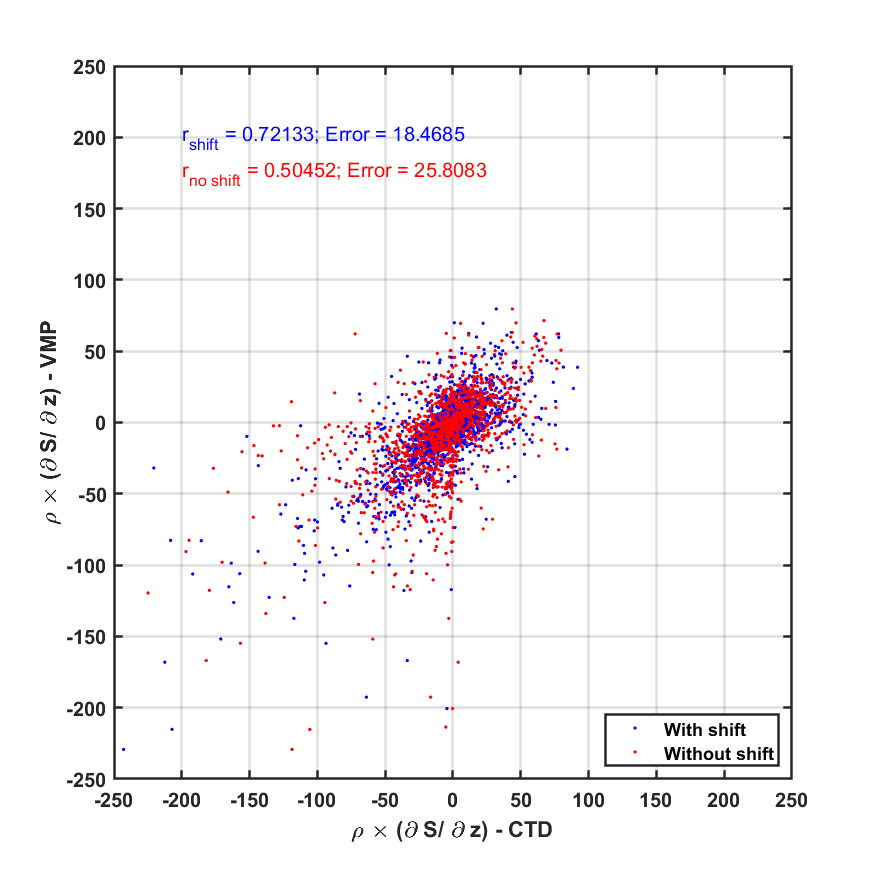}
	\centering
	\caption{Comparison of salinity gradients from CTD (x-axis) and VMP (y-axis). Blue dots show points from shifted CTD profiles, and red dots show non-shifted profiles. The text on the top left shows each scatter's correlation coefficient and error. We understand that shifting operation improves the match of the gradients from CTD and VMP.}
	\label{DO1p1}
\end{figure}

\begin{figure}[ht!]
	\includegraphics[width=1.1\textwidth]{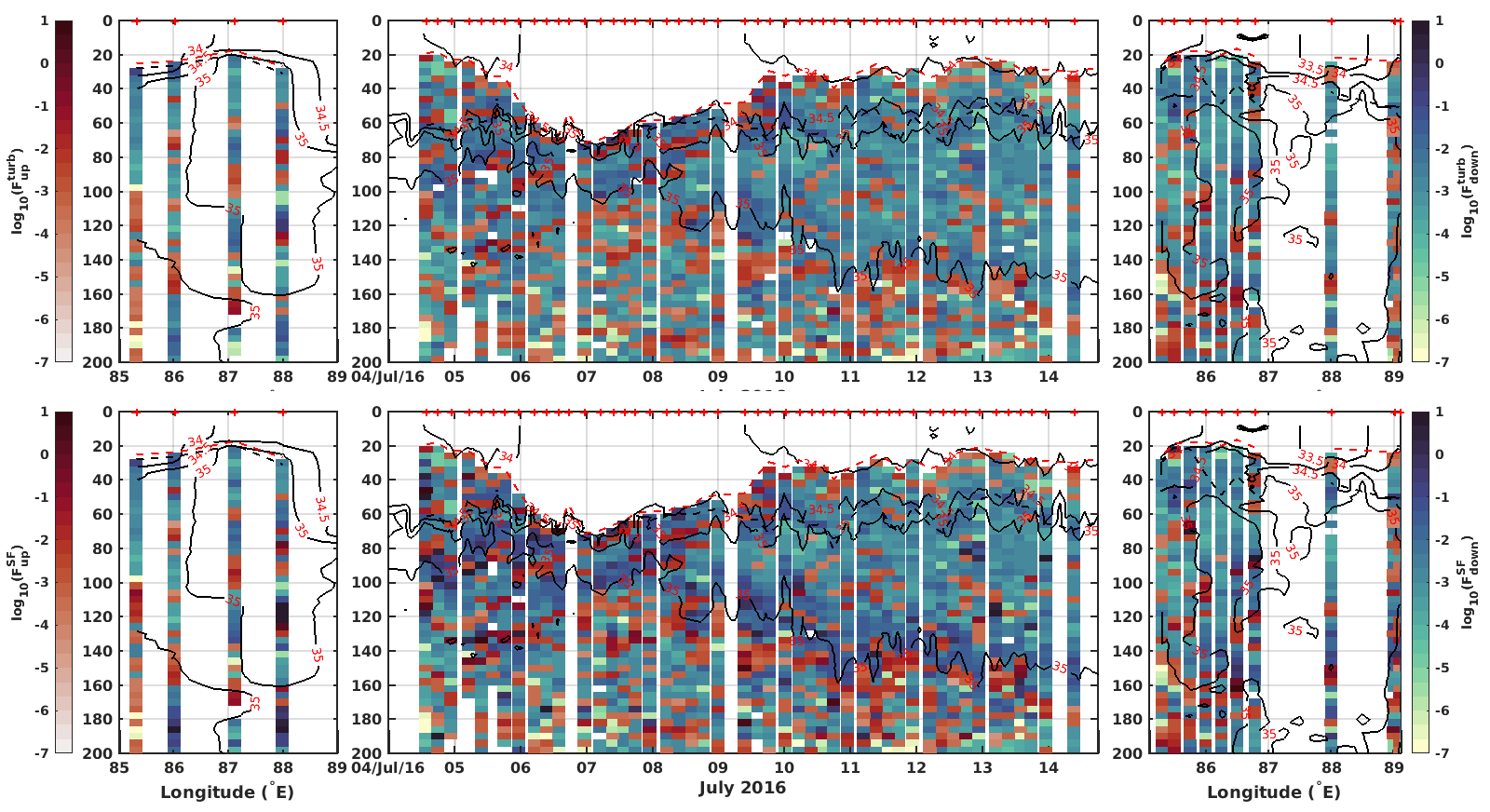}
	\centering
	\caption{The top row shows fluxes due to turbulent processes, and the bottom row shows fluxes when considering salt-fingering regions. The fluxes due to salt fingering are higher by an order around the bottom of HSC. Black lines show salinity contours. Dashed red and black lines are MLD and ILD, respectively. Fluxes shown in shades of blue are directed downwards, and shades of red are directed upwards. Compared to the diffusivity values, there are times and locations with missing flux profiles. These times and locations did not have simultaneous DO and microstructure profiles. Units of flux - $\upmu$mol m$^{-2}$ s$^{-1}$.}
	\label{DO5}
\end{figure}